\documentclass[aps,prl,twocolumn,showpacs]{revtex4-1}  
\usepackage{graphicx}
\usepackage[dvips]{epsfig}
\usepackage{bm}
\usepackage{amssymb,amsmath}
\usepackage[dvips,colorlinks=true,citecolor=blue]{hyperref}

\usepackage{natbib}

\newcommand{\be}{\begin{equation}}
\newcommand{\ee}{\end{equation}}
\newcommand{\ba}{\begin{eqnarray}}
\newcommand{\ea}{\end{eqnarray}}

\newcommand{\upa}{\uparrow}
\newcommand{\dna}{\downarrow}

\newcommand{\COMMENTED}[1]{}

\newcommand{\ca}[2]{ c_{#1 #2}^\dagger }
\newcommand{\de}[2]{c_{#1 #2}}

\newcommand{\ob}[1]{\langle #1\rangle}

\newcommand{\bfi}{\mathbf{i}}
\newcommand{\bfj}{\mathbf{j}}
\newcommand{\bfk}{\mathbf{k}}

\newcommand{\bfq}{\mathbf{q}}
\newcommand{\bfr}{\mathbf{r}}

\begin{document}

\title{Quantum disordered phase near the Mott
transition in the staggered-flux \\Hubbard model on a square lattice}

\author{Chia-Chen Chang}
\author{Richard T. Scalettar}
\affiliation{Department of Physics, University of California, Davis, 95616}

\begin{abstract}
We investigate ground state properties of the half-filled staggered-flux 
Hubbard model on a square lattice. Energy gaps to charge and spin excitations
and magnetic as well as dimer orders are calculated as a function of 
interaction strength $U/t$ by means of constrained-path quantum Monte Carlo 
method.
It is found that the system is a semi-metal at $U/t\lesssim 5.6$ and a Mott 
insulator with long-range antiferromagnetic order at $U/t \gtrsim 6.6$. 
In the range $5.6\lesssim U/t\lesssim 6.6$, the ground state is an correlated
insulator where both magnetic and dimer orders are absent. Furthermore, 
spin excitation in the intermediate phase appears to be gapless, and the
measured spin-spin correlation function exhibits power-law decaying behavior.
The data suggest that the non-magnetic ground state is a possible candidate
for the putative algebraic spin liquid.
\end{abstract}


\maketitle


At sufficiently low temperatures, condensed matter systems have a tendency to 
undergo phase transitions and develop long range order which reflects broken 
symmetry\cite{Landau:1937ta}. In a two-dimensional antiferromagnet,
however, Anderson recognized that the system could have a ground state that 
avoids all spontaneous symmetry-breaking and does not have magnetic order even 
at zero temperature\cite{Anderson:1973tm}.
Anderson's discovery, in conjunction with many subsequent theoretical investigations,
uncovered a new class of matter, named spin liquids, that go beyond Landau's paradigm.
Most notably, in contrast to conventional symmetry-breaking, spin liquids 
possess topological orders that cannot be characterized by local order parameters
and carry fractionalized excitations\cite{Wen:2002vk}.

Model Hamiltonians have played an important role in realizing such exotic spin liquid
states\cite{Marston:1991vv,Sachdev:1992ud}. 
Evidence of spin liquid phases has been found in the spin $1/2$ Heisenberg model on triangular
lattices\cite{Yunoki:2006ug}, square lattices with frustration\cite{Sandvik:2007dt,Jiang:2011ut,Wang:2011tv}, 
and kagome lattices\cite{Yan:2011vw}. 
In these geometrically frustrated systems\cite{Balents:2010ve}, antiferromagnetic (AF) 
orders are suppressed by strong quantum fluctuations. 
In addition to spin systems, there is also progress using the Hubbard model which contains 
spin and charge degrees of freedom. Spin liquid ground states have been identified in the
model on anisotropic triangular lattices\cite{Morita:2002ip} and on bipartite honeycomb 
lattices\cite{Meng:2010tf}.
%

In this paper, we examine ground state properties of the half-filled 
staggered-flux Hubbard model (sfHM) on a square lattice. As will be seen later,
low energy physics in the sfHM is described by Dirac fermions, similar to that
found in the Hubbard model on honeycomb lattices\cite{Meng:2010tf}.
The model is defined by the Hamiltonian
\ba
  H &=& -\sum_{\ob{\bfi\bfj},\sigma}( t_{\bfi\bfj}\ca{\bfi}{\sigma}\de{\bfj}{\sigma} 
     + t_{\bfj\bfi}\ca{\bfj}{\sigma}\de{\bfi}{\sigma} ) \nonumber\\
     && + U\sum_\bfi\,\left( n_{\bfi\upa}-\frac 1 2 \right)\left( n_{\bfi\dna}-\frac 1 2 \right),
\ea
where $t_{\bfi\bfj}=t\,e^{i\theta_{\bfi\bfj}}$ is the nearest-neighbor hopping and we
set $t=1$ throughout this work. The operator $c_{{\bfi}\sigma}^{\phantom{\dagger}}$ ($c_{\bfi\sigma}$)
creates (annihilates) an electron with spin $\sigma=\upa,\dna$ at site $\bfi$
on a lattice of size $N=L\times L$. $U>0$ is the onsite Coulomb repulsion.
We work in the canonical ensemble.

\begin{center}
\begin{figure}
\includegraphics[scale=0.34]{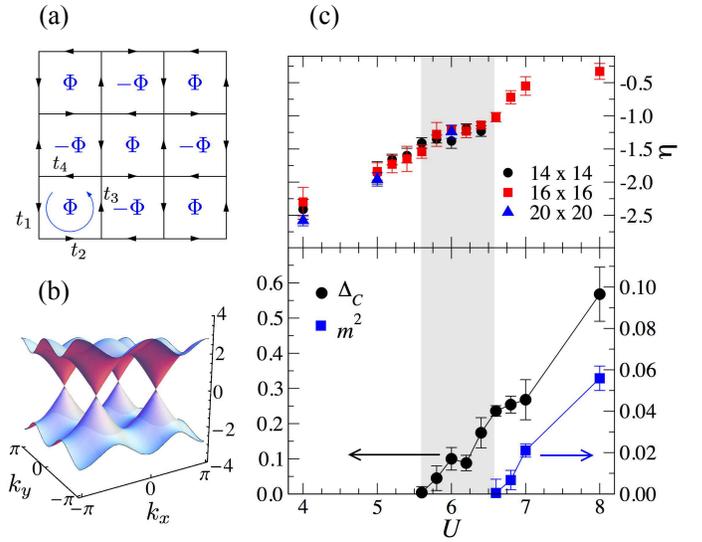}
\caption{(a) Arrangement of hopping amplitudes on a square lattice.
The phase $\Phi=\pi$ is distributed equally so that $t_1=t_2=t_3=t_4=e^{i\Phi/4}$.
(b) Band structure of the tight-binding Hamiltonian.
(c) Top panel: power law exponent $\eta$ extracted from fits to
staggered spin-spin correlation functions. 
Bottom panel: thermodynamic limit charge gap $\Delta_C$ and 
magnetic moment $m^2$ (see text for definition). The shaded region
indicates the non-magnetic insulating phase.}
\label{fig:phasediagram}
\end{figure}
\end{center}

An electron gains a phase $\Phi=\sum_\Box \theta_{\bfi\bfj}$ when it hops around a 
plaquette of the square lattice. $\Phi=0$ corresponds to the original Hubbard model.
We focus on the case $\Phi=\pi$ in the present study.
%
%
There is a gauge freedom in choosing $\theta_{\bfi\bfj}$. Here we distribute 
the phase $\Phi$ equally over all bonds around a plaquette and arrange the hoppings
according to Fig.~\ref{fig:phasediagram}(a). This leads to a lattice with 
plaquettes threaded alternatively by flux $\Phi$ and $-\Phi$. 
At $U=0$, the energy spectrum is $\epsilon_\bfk = \pm 2 \sqrt{ \cos^2 k_x + \cos^2 k_y}$.
The two energy bands meet at the Fermi surface $\epsilon_\bfk=0$ located at 
nodal points $\bfk_0=(\pm\pi/2,\pm\pi/2)$, as shown in Fig.~\ref{fig:phasediagram}(b). 
Close to the four nodal points the energy depends linearly on $\bfk$, which is 
similar to the massless Dirac spectrum found on the honeycomb lattice.

Our key result is that an intermediate non-magnetic insulating ground
state is identified between the semi-metal phase at weak interaction strengths and the
AF Mott insulator at strong couplings where the hopping terms become irrelevant. 
The calculated dimer correlation function shows that columnar valence bond order is also
absent in the intermediate phase. These results seem to indicate that the non-magnetic 
insulating phase is a candidate for the putative algebraic spin liquid ground state.
Therefore, our work suggests recent progress in optical lattice experiments\cite{Aidelsburger:2011hl} 
might provide a promising way of simulating the model, and observing this novel state
of matter.

\begin{figure}
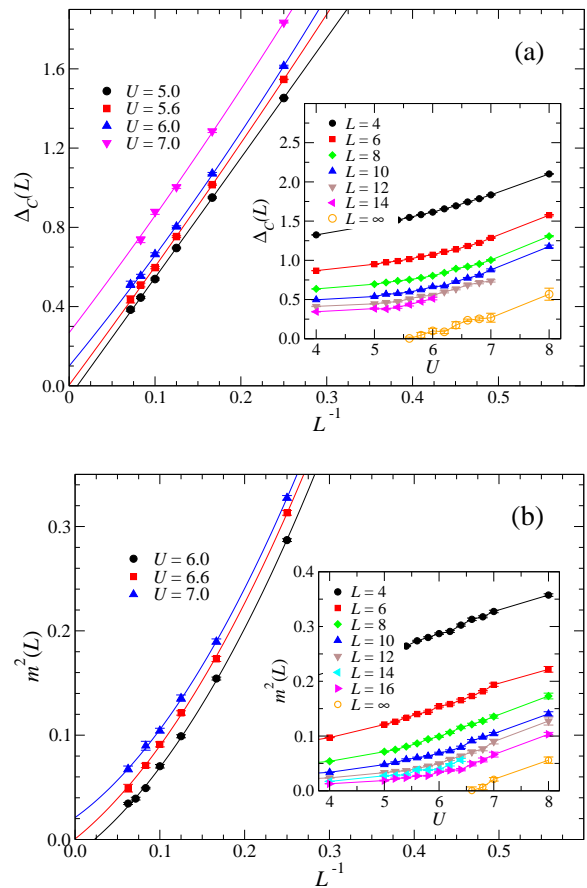

\includegraphics*[scale=0.29]{fig2a.eps}\\
\vspace{5mm}
\includegraphics*[scale=0.30]{fig2b.eps}
\caption{
(a) Extrapolation of the charge gap $\Delta_C(L)$.
Solid lines represent second-order polynomial fits to the QMC data.
Inset shows the charge gap $\Delta_C(L)$ as a function of $U$ obtained 
for $L=4$, 6, 8, 10, 12, 14, and extrapolated (empty circle) values. 
Lines are guides to the eyes.
(b) Finite size extrapolation of the spin structure factor $m^2(L)$.
Solid lines are second-order polynomial fits to the QMC data.
Inset: $m^2(L)$ versus $U$ on finite lattices and
extrapolated (empty circle) values. Lines are guide to the eyes.
}
\label{fig:charge.gap-Sk.spin}
\end{figure}

The sfHM is solved numerically by means of the constrained-path quantum Monte Carlo 
method\cite{Zhang:1995tl,*Zhang:1997ui}. Details of the method are described in the supplemental 
materials. We begin with the results for the charge excitation gap. 
In the canonical ensemble, the charge gap at half-filling can be defined as\cite{Furukawa:1992te}
\be
  \Delta_C(L) = \frac{1}{2}
     \left[
        E_g\left(\frac{N}{2}-1,\frac{N}{2}-1\right) - E_g\left(\frac{N}{2},\frac{N}{2}\right)
     \right],
\ee
which is the energy cost of removing a pair of electrons from the half-filled
ground state while keeping the system in the $S^z_{tot}=0$ sector.
We measure $\Delta_C(L)$ as a function of interaction strength $U$ on lattices with 
linear dimension up to $L=14$. As shown in the inset of Fig.~\ref{fig:charge.gap-Sk.spin}(a), 
the charge gap increases with $U$ on finite lattices. To pinpoint the critical interaction 
strength where the system turns into an insulator, we extrapolate $\Delta_C(L)$ at fixed 
$U$ to the thermodynamic limit $L\rightarrow\infty$ using second order polynomials in 
$L^{-1}$. The results, shown in the inset of Fig.~\ref{fig:charge.gap-Sk.spin}(a), indicate 
that the system is gapped for $U\gtrsim 5.6$. 

In addition to the charge excitation gap, AF long-range order is another 
essential feature characterizing a Mott insulator. To investigate whether there is any AF 
order in the ground state, we calculate the spin structure factor at the N{\'e}el wave 
vector $\bfq_{AF}=(\pi,\pi)$
\be
  S(\bfq_{AF},L) = \sum_\bfr\,e^{i\bfq_{AF}\cdot\bfr}\,\ob{S^x_\bfr S^x_0 + S^y_\bfr S^y_0 + S^z_\bfr S^z_0},
  \label{eq:SkSpin}
\ee
where $S^\delta_\bfr$ is the spin operator along the $\delta$-direction ($\delta=x,y,z$),
and $\ob{S^\delta_\bfr S^\delta_0}$ is the equal-time spin-spin correlation function.
Defining $m^2(L)=S(\bfq_{AF},L)/L^2$, a magnetically ordered phase is singled by a finite $m^2(L)$ 
in the thermodynamic limit.
The inset of Fig.~\ref{fig:charge.gap-Sk.spin}(b) shows the results of $m^2(L)$ as a function 
of $U$ on finite lattices. We use second-order polynomials in $L^{-1}$
to fit the QMC data and extract the value of $m^2(L)$ in the $L\rightarrow\infty$ limit.
It can be seen from the inset of Fig.~\ref{fig:charge.gap-Sk.spin}(b) that AF order kicks in 
at $U\gtrsim 6.6$, below which the system is in a paramagnetic phase.


Our analysis of charge gap and magnetic order above suggests that the ground state of 
the sfHM is a semi-metal at $U\lesssim 5.6$, and becomes a Mott insulator with long-range 
AF order at $U\gtrsim 6.6$. Therefore, unlike the original half-filled Hubbard model which 
enters the Mott phase at arbitrarily small $U$, the sfHM has a finite Mott transition point. 
This finding is consistent with a previous finite-temperature determinant quantum Monte 
Carlo work\cite{Otsuka:2002uj} which reports that the critical point of Mott transition 
in the staggered-flux model lies in the range $4\leq U\leq 8$. Moreover, our results indicate 
that in the region $5.6\lesssim U\lesssim 6.6$ there is an intermediate phase that is neither 
a semi-metal nor a Mott insulator.

\begin{figure}
\includegraphics*[scale=0.30]{fig3a.eps}\\
\vspace{3mm}
\includegraphics*[scale=0.29]{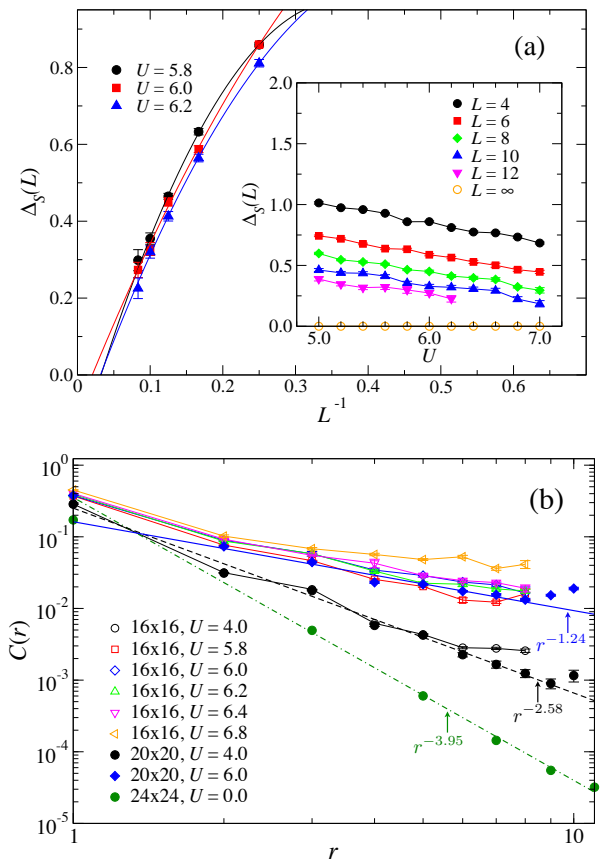}
\caption{(a) Finite size extrapolation of the spin gap $\Delta_S(L)$.
Solid lines are second-order polynomial fits to the Monte Carlo data.
The inset illustrates $\Delta_S(L)$ and its extrapolated value. 
Lines are guide to the eyes.
(b) Long-range behavior of the staggered spin-spin correlation function 
$C(\bfr)=(-1)^\bfr\ob{S^x_\bfr S^x_0 + S^y_\bfr S^y_0 + S^z_\bfr S^z_0}$
obtained on $L=16$, 20, and 24 ($U=0$ only). Straight lines are
representative power-law fits to the data: Green (dot-dashed) line:
$L=24$ at $U=0$, black (dashed) line: $L=20$ at $U=4$, and purple
(solid) line: $L=20$ at $U=6$.
}
\label{fig:spingap}
\end{figure}

The absence of AF order in the intermediate phase indicates that the ground 
state is dominated by short-range spin correlations. At large distances, the
spin correlation function could either decay exponentially or follow a
power-law. To study the nature of the non-magnetic insulating phase, we first 
calculate the spin excitation gap. Following Ref.~\cite{Noack:1995vd}, we 
write the spin gap at half-filling as
\be
  \Delta_S(L) = E_g\left(\frac{N}{2}+1,\frac{N}{2}-1\right) - E_g\left(\frac{N}{2},\frac{N}{2}\right),
\ee
which measures the energy cost of flipping an electron from spin-down to spin-up. 
Based on confinement arguments, a gapped spin excitation implies a
finite correlation length, leading to an exponentially decaying spin-spin correlation.
On the other hand, the correlation function would be described by a power-law if 
the spin excitation is gapless. We compute $\Delta_S(L)$ as a function of $U$ on finite
lattices. The spin gap results are shown in the inset of Fig.~\ref{fig:spingap}(a) for 
$5\leq U\leq 7$. The data at a given $U$ is then extrapolated to $L\rightarrow\infty$ 
using a second-order polynomial in $L^{-1}$ to extract the spin gap in the thermodynamic limit. 
Typical behavior of the fits is plotted in Fig.~\ref{fig:spingap}(a). As expected, the 
extrapolated spin gap remains zero in the gapless semi-metal phase ($U\lesssim 5.6$) and 
in the Mott phase ($U\gtrsim 6.6$) due to the presence of gapless spin wave excitations. 
More importantly, $\Delta_S(L)$ also shows gapless behavior in the region 
$5.6\lesssim U\lesssim 6.6$, implying that the spin-spin correlation should follow a 
power-law at large distances.

To support this observation, we plot in Fig.~\ref{fig:spingap}(b) the staggered
spin-spin correlation function
along the  $x$-axis. It appears that $C(\bfr)$ indeed decays algebraically
at large separations. Moreover, the correlation function decays more slowly with 
increasing $U$, and starts showing saturation in the Mott phase $U\gtrsim 6.6$.
In order to quantify the long-range behavior of $C(\bfr)$, we fit the staggered spin
correlation function to a power law $\alpha|\bfr|^\eta$ for $|\bfr|\geq 2$, where
$\alpha$ and $\eta$ are two fitting parameters. At $U=0$, it is known that $C(\bfr)$ 
decays as $|\bfr|^{-4}$\cite{Hermele:2005da}. This is also demonstrated in 
Fig.~\ref{fig:spingap}(b): the fitted exponent of $C(\bfr)$ for free fermions on a 
half-filled $24\times 24$ is $\eta=-3.95\pm0.13$. 
The exponent $\eta$ as a function of $U$ extracted from several half-filled lattices
is plotted in the top panel of Fig.~\ref{fig:phasediagram}(c). It can be seen from the
figure that $\eta$ immediately increases with $U$ from its non-interacting value due
to the effect of interaction. Although the data is quite scattered, the figure suggests 
that the exponent $\eta$ increases slowly with $U$ in the region $5.6\lesssim U\lesssim 6.6$.

\begin{figure}
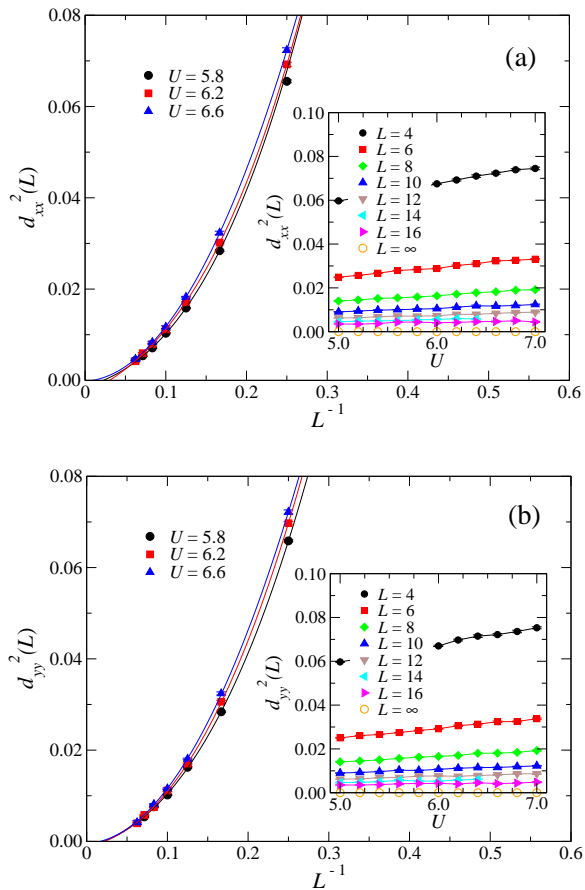

\includegraphics*[scale=0.30]{fig4a.eps}\\
\vspace{5mm}
\includegraphics*[scale=0.30]{fig4b.eps}
\caption{Thermodynamic limit extrapolation of the columnar VBS order 
In both figures, solid lines are three representative third-order polynomial 
fits to the Monte Carlo data. Insets shows the normalized dimer structure for 
single bonds in $x$ or $y$ directions on finite size lattices. Lines are guide 
to the eyes. The extrapolated value in the thermodynamic limit is indicated by 
orange (empty) circles in the insets.}
\label{fig:sk.dimer}
\end{figure}

Next we consider other order parameters proposed in Ref.~\cite{Hermele:2005da}.
The simplest scenario is the columnar valence bond solid (VBS) which breaks
translational symmetry. The VBS order can be probed by measuring the 
dimer structure factor 
\be
  D_{\delta\delta}(\bfq,L) =
  \frac 1 N \sum_\bfr\,e^{i\bfq\cdot\bfr}\,C^z_{\delta\delta}(\bfr),
\ee
where $C^z_{\delta\delta}(\bfr)$ is the $z$-component equal-time dimer-dimer correlation 
function for singlet bonds along the $\delta$-direction ($\delta=x,y$)
\be
  C^z_{\delta\delta}(\bfr)
  = \ob{ S^z_{\bfr+\hat\delta} S^z_\bfr S^z_{\hat\delta} S^z_0 } 
       -\ob{S^z_{\hat\delta} S^z_0}^2.
\ee
In the columnar VBS state, dimers line up coherently. Therefore 
$D_{\delta\delta}(\bfq,L)$ would pick up a characteristic momentum at $\bfk_{xx}=(\pi,0)$ 
or $\bfk_{yy}=(0,\pi)$ for $\delta=x$ or $y$ depending on the orientation of the bonds.
Indeed $D_{\delta\delta}(\bfq,L)$ peaks at $\bfk_{\delta\delta}$ in our finite size
simulations, as shown in the supplemental materials. To extract the VBS order in the 
thermodynamic limit, we calculate $d_{\delta\delta}^2(L)=D_{\delta\delta}(\bfk_{\delta\delta})/N$ 
and extrapolate to the $L\rightarrow\infty$ limit. As shown in Fig.~\ref{fig:sk.dimer}(a) 
and (b), both quantities vanish in the thermodynamic limit, implying the absence of 
columnar VBS order in the intermediate phase.


A commonly adopted definition of a spin liquid is that it is a non-magnetic Mott 
insulator in which neither spin nor lattice symmetry is broken. Based on this
definition, our numerical data presented in this work seems to suggest an {\it algebraic}
spin liquid ground state in the half-filled sfHM. However, the most unambiguous 
evidence of a spin liquid is its fractionalized excitation\cite{Misguich:2010vd}. 
Due to the nature of our method, we are not able to directly measure quantum properties
of excited states.  
In terms of the method, we note that although the half-filled staggered-flux model does not
have the fermion sign problem, we deliberately keep the constrained-path approximation and
calculate the ground state properties at half-filling. Our benchmark data shows that
the error appears to be small when compared with exact answers, as shown by the benchmark data
in supplemental materials. However, it is possible that the systematic error grows with $L$. 
A recent exact QMC method, linearized auxiliary fields Monte Carlo technique, reports that 
the half-filled ground state energy at $U=4$ is is $-0.85996(5)$\cite{Sorella:2011kd} per site 
in the thermodynamic limit. Our method, after boundary condition averaging, gives 
$-0.8559(4)$\cite{Chang:2008vg}, corresponding to a $0.47\%$ error. By including this systematic 
error, we estimate that the lower critical point where charge gap opens would be pushed to 
$U\sim 5.4\pm 0.1$.
In observable results such as correlation functions, extensive 
tests\cite{Zhang:1995tl,*Zhang:1997ui,Zhang:1997te} show that our systematic error is small
even at half-filling\cite{Zhang:1997te} and does not affect the physics of the 
numerical solutions.


To summarize, we have studied ground state properties in the half-filled staggered-flux 
Hubbard model on a square lattice. Charge and spin excitation gaps as well as spin and 
dimer orders are extracted by means of the constrained-path quantum Monte Carlo method. 
The system is found to be a semi-metal at $U\lesssim 5.6$ and an AF Mott insulator at
$U\gtrsim 6.6$. In the region $5.6\lesssim U\lesssim 6.6$, our data suggests that both
AF and VBS orders are absent in the ground state. Spin excitation in this region is 
gapless, a result that is consistent with the calculated staggered spin-spin correlation 
function which shows power-law decaying behavior at large distances.


We thank Professors Rajiv R. P. Singh and Shiwei Zhang for helpful comments. 
C.C. would like to thank Professor Hong Yao for illuminating
discussions. This research was supported by the DOE SciDAC (Grant No. DOE-DE-FC0206ER25793) 
and NSF PIF (Grant No. NSF-PHY-1005503) Programs. Computations were performed on Kraken at the 
National Institute for Computational Sciences.

\bibliography{fluxphase}

\begin{thebibliography}{24}%
\makeatletter
\providecommand \@ifxundefined [1]{%
 \@ifx{#1\undefined}
}%
\providecommand \@ifnum [1]{%
 \ifnum #1\expandafter \@firstoftwo
 \else \expandafter \@secondoftwo
 \fi
}%
\providecommand \@ifx [1]{%
 \ifx #1\expandafter \@firstoftwo
 \else \expandafter \@secondoftwo
 \fi
}%
\providecommand \natexlab [1]{#1}%
\providecommand \enquote  [1]{``#1''}%
\providecommand \bibnamefont  [1]{#1}%
\providecommand \bibfnamefont [1]{#1}%
\providecommand \citenamefont [1]{#1}%
\providecommand \href@noop [0]{\@secondoftwo}%
\providecommand \href [0]{\begingroup \@sanitize@url \@href}%
\providecommand \@href[1]{\@@startlink{#1}\@@href}%
\providecommand \@@href[1]{\endgroup#1\@@endlink}%
\providecommand \@sanitize@url [0]{\catcode `\\12\catcode `\$12\catcode
  `\&12\catcode `\#12\catcode `\^12\catcode `\_12\catcode `\%12\relax}%
\providecommand \@@startlink[1]{}%
\providecommand \@@endlink[0]{}%
\providecommand \url  [0]{\begingroup\@sanitize@url \@url }%
\providecommand \@url [1]{\endgroup\@href {#1}{\urlprefix }}%
\providecommand \urlprefix  [0]{URL }%
\providecommand \Eprint [0]{\href }%
\providecommand \doibase [0]{http://dx.doi.org/}%
\providecommand \selectlanguage [0]{\@gobble}%
\providecommand \bibinfo  [0]{\@secondoftwo}%
\providecommand \bibfield  [0]{\@secondoftwo}%
\providecommand \translation [1]{[#1]}%
\providecommand \BibitemOpen [0]{}%
\providecommand \bibitemStop [0]{}%
\providecommand \bibitemNoStop [0]{.\EOS\space}%
\providecommand \EOS [0]{\spacefactor3000\relax}%
\providecommand \BibitemShut  [1]{\csname bibitem#1\endcsname}%
\let\auto@bib@innerbib\@empty
\bibitem [{\citenamefont {Landau}(1937)}]{Landau:1937ta}%
  \BibitemOpen
  \bibfield  {author} {\bibinfo {author} {\bibfnamefont {L.}~\bibnamefont
  {Landau}},\ }\href@noop {} {\bibfield  {journal} {\bibinfo  {journal} {Phys.
  Z. Sowjun.}\ }\textbf {\bibinfo {volume} {11}},\ \bibinfo {pages} {26}
  (\bibinfo {year} {1937})}\BibitemShut {NoStop}%
\bibitem [{\citenamefont {Anderson}(1973)}]{Anderson:1973tm}%
  \BibitemOpen
  \bibfield  {author} {\bibinfo {author} {\bibfnamefont {P.}~\bibnamefont
  {Anderson}},\ }\href@noop {} {\bibfield  {journal} {\bibinfo  {journal} {Mat.
  Res. Bull.}\ }\textbf {\bibinfo {volume} {8}},\ \bibinfo {pages} {153}
  (\bibinfo {year} {1973})}\BibitemShut {NoStop}%
\bibitem [{\citenamefont {Wen}(2002)}]{Wen:2002vk}%
  \BibitemOpen
  \bibfield  {author} {\bibinfo {author} {\bibfnamefont {X.-G.}\ \bibnamefont
  {Wen}},\ }\href@noop {} {\bibfield  {journal} {\bibinfo  {journal} {Phys.
  Rev. B}\ }\textbf {\bibinfo {volume} {65}},\ \bibinfo {pages} {165113}
  (\bibinfo {year} {2002})}\BibitemShut {NoStop}%
\bibitem [{\citenamefont {Marston}\ and\ \citenamefont
  {Zeng}(1991)}]{Marston:1991vv}%
  \BibitemOpen
  \bibfield  {author} {\bibinfo {author} {\bibfnamefont {J.}~\bibnamefont
  {Marston}}\ and\ \bibinfo {author} {\bibfnamefont {C.}~\bibnamefont {Zeng}},\
  }\href@noop {} {\bibfield  {journal} {\bibinfo  {journal} {J. App. Phys.}\
  }\textbf {\bibinfo {volume} {69}},\ \bibinfo {pages} {5962} (\bibinfo {year}
  {1991})}\BibitemShut {NoStop}%
\bibitem [{\citenamefont {Sachdev}(1992)}]{Sachdev:1992ud}%
  \BibitemOpen
  \bibfield  {author} {\bibinfo {author} {\bibfnamefont {S.}~\bibnamefont
  {Sachdev}},\ }\href@noop {} {\bibfield  {journal} {\bibinfo  {journal} {Phys.
  Rev. B}\ }\textbf {\bibinfo {volume} {45}},\ \bibinfo {pages} {12377}
  (\bibinfo {year} {1992})}\BibitemShut {NoStop}%
\bibitem [{\citenamefont {Yunoki}\ and\ \citenamefont
  {Sorella}(2006)}]{Yunoki:2006ug}%
  \BibitemOpen
  \bibfield  {author} {\bibinfo {author} {\bibfnamefont {S.}~\bibnamefont
  {Yunoki}}\ and\ \bibinfo {author} {\bibfnamefont {S.}~\bibnamefont
  {Sorella}},\ }\href@noop {} {\bibfield  {journal} {\bibinfo  {journal} {Phys.
  Rev. B}\ }\textbf {\bibinfo {volume} {74}} (\bibinfo {year}
  {2006})}\BibitemShut {NoStop}%
\bibitem [{\citenamefont {Sandvik}(2007)}]{Sandvik:2007dt}%
  \BibitemOpen
  \bibfield  {author} {\bibinfo {author} {\bibfnamefont {A.}~\bibnamefont
  {Sandvik}},\ }\href@noop {} {\bibfield  {journal} {\bibinfo  {journal} {Phys.
  Rev. Lett.}\ }\textbf {\bibinfo {volume} {98}} (\bibinfo {year}
  {2007})}\BibitemShut {NoStop}%
\bibitem [{\citenamefont {Jiang}\ \emph {et~al.}(2011)\citenamefont {Jiang},
  \citenamefont {Yao},\ and\ \citenamefont {Balents}}]{Jiang:2011ut}%
  \BibitemOpen
  \bibfield  {author} {\bibinfo {author} {\bibfnamefont {H.-C.}\ \bibnamefont
  {Jiang}}, \bibinfo {author} {\bibfnamefont {H.}~\bibnamefont {Yao}}, \ and\
  \bibinfo {author} {\bibfnamefont {L.}~\bibnamefont {Balents}},\ }\href@noop
  {} {\bibfield  {journal} {\bibinfo  {journal} {arXiv:1112.2241}\ } (\bibinfo
  {year} {2011})}\BibitemShut {NoStop}%
\bibitem [{\citenamefont {Wang}\ \emph {et~al.}(2011)\citenamefont {Wang},
  \citenamefont {Gu}, \citenamefont {Verstraete},\ and\ \citenamefont
  {Wen}}]{Wang:2011tv}%
  \BibitemOpen
  \bibfield  {author} {\bibinfo {author} {\bibfnamefont {L.}~\bibnamefont
  {Wang}}, \bibinfo {author} {\bibfnamefont {Z.-C.}\ \bibnamefont {Gu}},
  \bibinfo {author} {\bibfnamefont {F.}~\bibnamefont {Verstraete}}, \ and\
  \bibinfo {author} {\bibfnamefont {X.-G.}\ \bibnamefont {Wen}},\ }\href@noop
  {} {\bibfield  {journal} {\bibinfo  {journal} {arXiv:1112.3331}\ } (\bibinfo
  {year} {2011})}\BibitemShut {NoStop}%
\bibitem [{\citenamefont {Yan}\ \emph {et~al.}(2011)\citenamefont {Yan},
  \citenamefont {Huse},\ and\ \citenamefont {White}}]{Yan:2011vw}%
  \BibitemOpen
  \bibfield  {author} {\bibinfo {author} {\bibfnamefont {S.}~\bibnamefont
  {Yan}}, \bibinfo {author} {\bibfnamefont {D.~A.}\ \bibnamefont {Huse}}, \
  and\ \bibinfo {author} {\bibfnamefont {S.~R.}\ \bibnamefont {White}},\
  }\href@noop {} {\bibfield  {journal} {\bibinfo  {journal} {Science}\ }\textbf
  {\bibinfo {volume} {332}},\ \bibinfo {pages} {1173} (\bibinfo {year}
  {2011})}\BibitemShut {NoStop}%
\bibitem [{\citenamefont {Balents}(2010)}]{Balents:2010ve}%
  \BibitemOpen
  \bibfield  {author} {\bibinfo {author} {\bibfnamefont {L.}~\bibnamefont
  {Balents}},\ }\href@noop {} {\bibfield  {journal} {\bibinfo  {journal}
  {Nature}\ }\textbf {\bibinfo {volume} {464}},\ \bibinfo {pages} {199}
  (\bibinfo {year} {2010})}\BibitemShut {NoStop}%
\bibitem [{\citenamefont {Morita}\ \emph {et~al.}(2002)\citenamefont {Morita},
  \citenamefont {Watanabe},\ and\ \citenamefont {Imada}}]{Morita:2002ip}%
  \BibitemOpen
  \bibfield  {author} {\bibinfo {author} {\bibfnamefont {H.}~\bibnamefont
  {Morita}}, \bibinfo {author} {\bibfnamefont {S.}~\bibnamefont {Watanabe}}, \
  and\ \bibinfo {author} {\bibfnamefont {M.}~\bibnamefont {Imada}},\
  }\href@noop {} {\bibfield  {journal} {\bibinfo  {journal} {J. Phys. Soc.
  Jpn.}\ }\textbf {\bibinfo {volume} {71}},\ \bibinfo {pages} {2109} (\bibinfo
  {year} {2002})}\BibitemShut {NoStop}%
\bibitem [{\citenamefont {Meng}\ \emph {et~al.}(2010)\citenamefont {Meng},
  \citenamefont {Lang}, \citenamefont {Wessel}, \citenamefont {Assaad},\ and\
  \citenamefont {Muramatsu}}]{Meng:2010tf}%
  \BibitemOpen
  \bibfield  {author} {\bibinfo {author} {\bibfnamefont {Z.}~\bibnamefont
  {Meng}}, \bibinfo {author} {\bibfnamefont {T.}~\bibnamefont {Lang}}, \bibinfo
  {author} {\bibfnamefont {S.}~\bibnamefont {Wessel}}, \bibinfo {author}
  {\bibfnamefont {F.}~\bibnamefont {Assaad}}, \ and\ \bibinfo {author}
  {\bibfnamefont {A.}~\bibnamefont {Muramatsu}},\ }\href@noop {} {\bibfield
  {journal} {\bibinfo  {journal} {Nature}\ }\textbf {\bibinfo {volume} {464}},\
  \bibinfo {pages} {847} (\bibinfo {year} {2010})}\BibitemShut {NoStop}%
\bibitem [{\citenamefont {Aidelsburger}\ \emph {et~al.}(2011)\citenamefont
  {Aidelsburger}, \citenamefont {Atala}, \citenamefont {Nascimb{\`e}ne},
  \citenamefont {Trotzky}, \citenamefont {Chen},\ and\ \citenamefont
  {Bloch}}]{Aidelsburger:2011hl}%
  \BibitemOpen
  \bibfield  {author} {\bibinfo {author} {\bibfnamefont {M.}~\bibnamefont
  {Aidelsburger}}, \bibinfo {author} {\bibfnamefont {M.}~\bibnamefont {Atala}},
  \bibinfo {author} {\bibfnamefont {S.}~\bibnamefont {Nascimb{\`e}ne}},
  \bibinfo {author} {\bibfnamefont {S.}~\bibnamefont {Trotzky}}, \bibinfo
  {author} {\bibfnamefont {Y.~A.}\ \bibnamefont {Chen}}, \ and\ \bibinfo
  {author} {\bibfnamefont {I.}~\bibnamefont {Bloch}},\ }\href@noop {}
  {\bibfield  {journal} {\bibinfo  {journal} {Phys. Rev. Lett.}\ }\textbf
  {\bibinfo {volume} {107}},\ \bibinfo {pages} {255301} (\bibinfo {year}
  {2011})}\BibitemShut {NoStop}%
\bibitem [{\citenamefont {Zhang}\ \emph {et~al.}(1995)\citenamefont {Zhang},
  \citenamefont {Carlson},\ and\ \citenamefont {Gubernatis}}]{Zhang:1995tl}%
  \BibitemOpen
  \bibfield  {author} {\bibinfo {author} {\bibfnamefont {S.}~\bibnamefont
  {Zhang}}, \bibinfo {author} {\bibfnamefont {J.}~\bibnamefont {Carlson}}, \
  and\ \bibinfo {author} {\bibfnamefont {J.}~\bibnamefont {Gubernatis}},\
  }\href@noop {} {\bibfield  {journal} {\bibinfo  {journal} {Phys. Rev. Lett.}\
  }\textbf {\bibinfo {volume} {74}},\ \bibinfo {pages} {3652} (\bibinfo {year}
  {1995})}\BibitemShut {NoStop}%
\bibitem [{\citenamefont {Zhang}\ \emph
  {et~al.}(1997{\natexlab{a}})\citenamefont {Zhang}, \citenamefont {Carlson},\
  and\ \citenamefont {Gubernatis}}]{Zhang:1997ui}%
  \BibitemOpen
  \bibfield  {author} {\bibinfo {author} {\bibfnamefont {S.}~\bibnamefont
  {Zhang}}, \bibinfo {author} {\bibfnamefont {J.}~\bibnamefont {Carlson}}, \
  and\ \bibinfo {author} {\bibfnamefont {J.}~\bibnamefont {Gubernatis}},\
  }\href@noop {} {\bibfield  {journal} {\bibinfo  {journal} {Phys. Rev. B}\
  }\textbf {\bibinfo {volume} {55}},\ \bibinfo {pages} {7464} (\bibinfo {year}
  {1997}{\natexlab{a}})}\BibitemShut {NoStop}%
\bibitem [{\citenamefont {Furukawa}\ and\ \citenamefont
  {Imada}(1992)}]{Furukawa:1992te}%
  \BibitemOpen
  \bibfield  {author} {\bibinfo {author} {\bibfnamefont {N.}~\bibnamefont
  {Furukawa}}\ and\ \bibinfo {author} {\bibfnamefont {M.}~\bibnamefont
  {Imada}},\ }\href@noop {} {\bibfield  {journal} {\bibinfo  {journal} {J.
  Phys. Soc. Jpn.}\ }\textbf {\bibinfo {volume} {61}},\ \bibinfo {pages} {3331}
  (\bibinfo {year} {1992})}\BibitemShut {NoStop}%
\bibitem [{\citenamefont {Otsuka}\ and\ \citenamefont
  {Hatsugai}(2002)}]{Otsuka:2002uj}%
  \BibitemOpen
  \bibfield  {author} {\bibinfo {author} {\bibfnamefont {Y.}~\bibnamefont
  {Otsuka}}\ and\ \bibinfo {author} {\bibfnamefont {Y.}~\bibnamefont
  {Hatsugai}},\ }\href@noop {} {\bibfield  {journal} {\bibinfo  {journal}
  {Phys. Rev. B}\ }\textbf {\bibinfo {volume} {65}},\ \bibinfo {pages} {073101}
  (\bibinfo {year} {2002})}\BibitemShut {NoStop}%
\bibitem [{\citenamefont {Noack}\ \emph {et~al.}(1995)\citenamefont {Noack},
  \citenamefont {White},\ and\ \citenamefont {Scalapino}}]{Noack:1995vd}%
  \BibitemOpen
  \bibfield  {author} {\bibinfo {author} {\bibfnamefont {R.}~\bibnamefont
  {Noack}}, \bibinfo {author} {\bibfnamefont {S.}~\bibnamefont {White}}, \ and\
  \bibinfo {author} {\bibfnamefont {D.}~\bibnamefont {Scalapino}},\ }\href@noop
  {} {\bibfield  {journal} {\bibinfo  {journal} {Europhys. Lett.}\ }\textbf
  {\bibinfo {volume} {30}},\ \bibinfo {pages} {163} (\bibinfo {year}
  {1995})}\BibitemShut {NoStop}%
\bibitem [{\citenamefont {Hermele}\ \emph {et~al.}(2005)\citenamefont
  {Hermele}, \citenamefont {Senthil},\ and\ \citenamefont
  {Fisher}}]{Hermele:2005da}%
  \BibitemOpen
  \bibfield  {author} {\bibinfo {author} {\bibfnamefont {M.}~\bibnamefont
  {Hermele}}, \bibinfo {author} {\bibfnamefont {T.}~\bibnamefont {Senthil}}, \
  and\ \bibinfo {author} {\bibfnamefont {M.~P.~A.}\ \bibnamefont {Fisher}},\
  }\href@noop {} {\bibfield  {journal} {\bibinfo  {journal} {Phys. Rev. B}\
  }\textbf {\bibinfo {volume} {72}} (\bibinfo {year} {2005})}\BibitemShut
  {NoStop}%
\bibitem [{\citenamefont {Misguich}(2010)}]{Misguich:2010vd}%
  \BibitemOpen
  \bibfield  {author} {\bibinfo {author} {\bibfnamefont {G.}~\bibnamefont
  {Misguich}},\ }in\ \href@noop {} {\emph {\bibinfo {booktitle} {Introduction
  to Frustrated Magnetism}}},\ \bibinfo {editor} {edited by\ \bibinfo {editor}
  {\bibfnamefont {C.}~\bibnamefont {Lacroix}}, \bibinfo {editor} {\bibfnamefont
  {F.}~\bibnamefont {Mila}}, \ and\ \bibinfo {editor} {\bibfnamefont
  {P.}~\bibnamefont {Mendels}}}\ (\bibinfo  {publisher} {Springer},\ \bibinfo
  {year} {2010})\ p.\ \bibinfo {pages} {407}\BibitemShut {NoStop}%
\bibitem [{\citenamefont {Sorella}(2011)}]{Sorella:2011kd}%
  \BibitemOpen
  \bibfield  {author} {\bibinfo {author} {\bibfnamefont {S.}~\bibnamefont
  {Sorella}},\ }\href@noop {} {\bibfield  {journal} {\bibinfo  {journal} {Phys.
  Rev. B}\ }\textbf {\bibinfo {volume} {84}},\ \bibinfo {pages} {241110(R)}
  (\bibinfo {year} {2011})}\BibitemShut {NoStop}%
\bibitem [{\citenamefont {Chang}\ and\ \citenamefont
  {Zhang}(2008)}]{Chang:2008vg}%
  \BibitemOpen
  \bibfield  {author} {\bibinfo {author} {\bibfnamefont {C.-C.}\ \bibnamefont
  {Chang}}\ and\ \bibinfo {author} {\bibfnamefont {S.}~\bibnamefont {Zhang}},\
  }\href@noop {} {\bibfield  {journal} {\bibinfo  {journal} {Phys. Rev. B}\
  }\textbf {\bibinfo {volume} {78}},\ \bibinfo {pages} {165101} (\bibinfo
  {year} {2008})}\BibitemShut {NoStop}%
\bibitem [{\citenamefont {Zhang}\ \emph
  {et~al.}(1997{\natexlab{b}})\citenamefont {Zhang}, \citenamefont {Carlson},\
  and\ \citenamefont {Gubernatis}}]{Zhang:1997te}%
  \BibitemOpen
  \bibfield  {author} {\bibinfo {author} {\bibfnamefont {S.}~\bibnamefont
  {Zhang}}, \bibinfo {author} {\bibfnamefont {J.}~\bibnamefont {Carlson}}, \
  and\ \bibinfo {author} {\bibfnamefont {J.}~\bibnamefont {Gubernatis}},\
  }\href@noop {} {\bibfield  {journal} {\bibinfo  {journal} {Phys. Rev. Lett.}\
  }\textbf {\bibinfo {volume} {78}},\ \bibinfo {pages} {4486} (\bibinfo {year}
  {1997}{\natexlab{b}})}\BibitemShut {NoStop}%
\end{thebibliography}%


\begin{thebibliography}{7}%
\makeatletter
\providecommand \@ifxundefined [1]{%
 \@ifx{#1\undefined}
}%
\providecommand \@ifnum [1]{%
 \ifnum #1\expandafter \@firstoftwo
 \else \expandafter \@secondoftwo
 \fi
}%
\providecommand \@ifx [1]{%
 \ifx #1\expandafter \@firstoftwo
 \else \expandafter \@secondoftwo
 \fi
}%
\providecommand \natexlab [1]{#1}%
\providecommand \enquote  [1]{``#1''}%
\providecommand \bibnamefont  [1]{#1}%
\providecommand \bibfnamefont [1]{#1}%
\providecommand \citenamefont [1]{#1}%
\providecommand \href@noop [0]{\@secondoftwo}%
\providecommand \href [0]{\begingroup \@sanitize@url \@href}%
\providecommand \@href[1]{\@@startlink{#1}\@@href}%
\providecommand \@@href[1]{\endgroup#1\@@endlink}%
\providecommand \@sanitize@url [0]{\catcode `\\12\catcode `\$12\catcode
  `\&12\catcode `\#12\catcode `\^12\catcode `\_12\catcode `\%12\relax}%
\providecommand \@@startlink[1]{}%
\providecommand \@@endlink[0]{}%
\providecommand \url  [0]{\begingroup\@sanitize@url \@url }%
\providecommand \@url [1]{\endgroup\@href {#1}{\urlprefix }}%
\providecommand \urlprefix  [0]{URL }%
\providecommand \Eprint [0]{\href }%
\providecommand \doibase [0]{http://dx.doi.org/}%
\providecommand \selectlanguage [0]{\@gobble}%
\providecommand \bibinfo  [0]{\@secondoftwo}%
\providecommand \bibfield  [0]{\@secondoftwo}%
\providecommand \translation [1]{[#1]}%
\providecommand \BibitemOpen [0]{}%
\providecommand \bibitemStop [0]{}%
\providecommand \bibitemNoStop [0]{.\EOS\space}%
\providecommand \EOS [0]{\spacefactor3000\relax}%
\providecommand \BibitemShut  [1]{\csname bibitem#1\endcsname}%
\let\auto@bib@innerbib\@empty
\bibitem [{\citenamefont {Zhang}\ \emph {et~al.}(1995)\citenamefont {Zhang},
  \citenamefont {Carlson},\ and\ \citenamefont {Gubernatis}}]{Zhang:1995tl}%
  \BibitemOpen
  \bibfield  {author} {\bibinfo {author} {\bibfnamefont {S.}~\bibnamefont
  {Zhang}}, \bibinfo {author} {\bibfnamefont {J.}~\bibnamefont {Carlson}}, \
  and\ \bibinfo {author} {\bibfnamefont {J.}~\bibnamefont {Gubernatis}},\
  }\href@noop {} {\bibfield  {journal} {\bibinfo  {journal} {Phys. Rev. Lett.}\
  }\textbf {\bibinfo {volume} {74}},\ \bibinfo {pages} {3652} (\bibinfo {year}
  {1995})}\BibitemShut {NoStop}%
\bibitem [{\citenamefont {Zhang}\ \emph
  {et~al.}(1997{\natexlab{a}})\citenamefont {Zhang}, \citenamefont {Carlson},\
  and\ \citenamefont {Gubernatis}}]{Zhang:1997ui}%
  \BibitemOpen
  \bibfield  {author} {\bibinfo {author} {\bibfnamefont {S.}~\bibnamefont
  {Zhang}}, \bibinfo {author} {\bibfnamefont {J.}~\bibnamefont {Carlson}}, \
  and\ \bibinfo {author} {\bibfnamefont {J.}~\bibnamefont {Gubernatis}},\
  }\href@noop {} {\bibfield  {journal} {\bibinfo  {journal} {Phys. Rev. B}\
  }\textbf {\bibinfo {volume} {55}},\ \bibinfo {pages} {7464} (\bibinfo {year}
  {1997}{\natexlab{a}})}\BibitemShut {NoStop}%
\bibitem [{\citenamefont {Suzuki}(1976)}]{Suzuki:1976vp}%
  \BibitemOpen
  \bibfield  {author} {\bibinfo {author} {\bibfnamefont {M.}~\bibnamefont
  {Suzuki}},\ }\href@noop {} {\bibfield  {journal} {\bibinfo  {journal} {Comm.
  Math. Phys.}\ }\textbf {\bibinfo {volume} {51}},\ \bibinfo {pages} {183}
  (\bibinfo {year} {1976})}\BibitemShut {NoStop}%
\bibitem [{\citenamefont {Hirsch}(1983)}]{Hirsch:1983uc}%
  \BibitemOpen
  \bibfield  {author} {\bibinfo {author} {\bibfnamefont {J.}~\bibnamefont
  {Hirsch}},\ }\href@noop {} {\bibfield  {journal} {\bibinfo  {journal} {Phys.
  Rev. B}\ }\textbf {\bibinfo {volume} {28}},\ \bibinfo {pages} {4059}
  (\bibinfo {year} {1983})}\BibitemShut {NoStop}%
\bibitem [{\citenamefont {Kalos}(1967)}]{Kalos:1967wv}%
  \BibitemOpen
  \bibfield  {author} {\bibinfo {author} {\bibfnamefont {M.}~\bibnamefont
  {Kalos}},\ }\href@noop {} {\bibfield  {journal} {\bibinfo  {journal} {J.
  Comput. Phys.}\ }\textbf {\bibinfo {volume} {2}},\ \bibinfo {pages} {257}
  (\bibinfo {year} {1967})}\BibitemShut {NoStop}%
\bibitem [{\citenamefont {Dagotto}\ \emph {et~al.}(1992)\citenamefont
  {Dagotto}, \citenamefont {Moreo}, \citenamefont {Ortolani}, \citenamefont
  {Poilblanc},\ and\ \citenamefont {Riera}}]{Dagotto:1992uf}%
  \BibitemOpen
  \bibfield  {author} {\bibinfo {author} {\bibfnamefont {E.}~\bibnamefont
  {Dagotto}}, \bibinfo {author} {\bibfnamefont {A.}~\bibnamefont {Moreo}},
  \bibinfo {author} {\bibfnamefont {F.}~\bibnamefont {Ortolani}}, \bibinfo
  {author} {\bibfnamefont {D.}~\bibnamefont {Poilblanc}}, \ and\ \bibinfo
  {author} {\bibfnamefont {J.}~\bibnamefont {Riera}},\ }\href@noop {}
  {\bibfield  {journal} {\bibinfo  {journal} {Phys. Rev. B}\ }\textbf {\bibinfo
  {volume} {45}} (\bibinfo {year} {1992})}\BibitemShut {NoStop}%
\bibitem [{\citenamefont {Zhang}\ \emph
  {et~al.}(1997{\natexlab{b}})\citenamefont {Zhang}, \citenamefont {Carlson},\
  and\ \citenamefont {Gubernatis}}]{Zhang:1997te}%
  \BibitemOpen
  \bibfield  {author} {\bibinfo {author} {\bibfnamefont {S.}~\bibnamefont
  {Zhang}}, \bibinfo {author} {\bibfnamefont {J.}~\bibnamefont {Carlson}}, \
  and\ \bibinfo {author} {\bibfnamefont {J.}~\bibnamefont {Gubernatis}},\
  }\href@noop {} {\bibfield  {journal} {\bibinfo  {journal} {Phys. Rev. Lett.}\
  }\textbf {\bibinfo {volume} {78}},\ \bibinfo {pages} {4486} (\bibinfo {year}
  {1997}{\natexlab{b}})}\BibitemShut {NoStop}%
\end{thebibliography}%

\end{document}


\title{Supplemental Material for ``Quantum disordered phase near the Mott
transition in the staggered-flux Hubbard model on a square lattice''
}

\author{Chia-Chen Chang}
\author{Richard T. Scalettar}
\affiliation{Department of Physics, University of California, Davis, 95616}


\maketitle


\section{Method and Benchmarks}

The square lattice staggered-flux Hubbard model (sfHM) is solved numerically
by using the ground state constrained-path quantum Monte Carlo (CPQMC) 
method\cite{Zhang:1995tl,*Zhang:1997ui}. The CPQMC method projects the many-body 
ground state wave function $\ket{\Psi_0}$ from a trial state $\ket{\Psi_T}$, 
assuming $\ob{\Psi_T|\Psi_0}\neq 0$, by successively applying an imaginary-time 
propagator $e^{-\Delta\tau H}$ to $\ket{\Psi_T}$ with $\Delta\tau$ being the 
imaginary-time step.
The propagator is decomposed according to the second order Trotter-Suzuki 
formula\cite{Suzuki:1976vp}. The two-body part of the resulting operator is 
then transformed into one-body projectors using a spin-decomposed Hubbard-Stratonovish 
(HS) transformation\cite{Hirsch:1983uc}.
Apart from systematic errors due to Trotter break-up, we arrive at a formally
exact expression $e^{-\Delta\tau H}=\sum_{\{\mathbf{x}\}}P(\{\mathbf{x}\})B(\{\mathbf{x}\})$,
where $\{\mathbf{x}\}$ is a collection of $N$ Ising-like HS variables,
$P(\{\mathbf{x}\})$ is their probability distribution, and
$B(\{\mathbf{x}\})$ is a one-body projector.
The projection is then realized by importance-sampled open-ended random walks 
with non-orthogonal Slater determinants (SDs), where the projectors 
$B(\{\mathbf{x}\})$ propagate one SD into another.

Away from half-filling (one electron per lattice site), 
the fermion sign problem is controlled by the constrained-path 
approximation\cite{Zhang:1995tl,*Zhang:1997ui}.
The ground state wave function obtained by the CPQMC method is written as 
$|\Psi_0^c\rangle = \sum_\phi w(\phi) |\phi\rangle$, where $|\phi\rangle$ are 
SDs sampled by the QMC, $w(\phi)$ are weight factors dictated by the
distribution of $\ket{\phi}$.
Since the Schr\"odinger equation is linear, $|\Psi_0^c\rangle$ and $-|\Psi_0^c\rangle$
are two degenerate solutions which can be both sampled in a random walk.
The appearance of the two sets with opposite signs in the Monte
Carlo samples is the origin of the exponential sign decay.
To control the problem, we restrict the walkers such that at each step
of projection the condition $\langle\Psi_T|\phi\rangle >0$ is fulfilled.

After the random walk has equilibrated, expectation values can be computed
from $\ket{\Psi_0^c}$. For example, the ground state energy is evaluated
using a mixed estimator
\be
  E_g = \frac{\ob{\Psi_T|H|\Psi_0^c}}{\ob{\Psi_T|\Psi_0^c}},
\ee
where $H$ is the Hamiltonian. For observables that does not commute with $H$,
we use a scheme called back-propagation (BP)\cite{Zhang:1997ui} which is
similar to the forward-walking technique in the Green's Function Monte Carlo
method\cite{Kalos:1967wv}. 
Throughout this work, we use free electron wave functions as our $\ket{\Psi_T}$.
A very small amount of anisotropy ($\delta\Phi=0.002\Phi$) is added to the total
flux per plaquette $\Phi=\pi$ in order to lift the two-fold degeneracy in the finite size 
single-particle spectrum. We also average our results over boundary conditions
so that finite size effects can be minimized.

To access the accuracy of our method, we compare the energy of a $4\times 4$
Hubbard cluster at $U=4$ and $\Phi=0$. 
Let $E_g(N_\upa,N_\dna)$ denote the energy of the system with $N_\upa$ ($N_\dna$) 
spin-up (-down) electrons. The exact energy per site is $E_g(7,7)/N=-0.9831$\cite{Dagotto:1992uf},
where $N=L\times L$ is the number of lattice sites of an $L\times L$ square lattice.
After correcting the Trotter error, the CPQMC method gives $-0.9821(6)$, 
corresponding to an error of $0.1\%$. At half-filling, the exact energy is
$E_g(8,8)/N=-0.8513$\cite{Dagotto:1992uf}. The CPQMC energy obtained {\it with} 
contrained-path approximation is $-0.8491(2)$, which is within $0.26\%$ of the exact 
energy. Comparisons of other observables, such as spin-spin correlation function, computed 
using the BP scheme can be found in Ref. \onlinecite{Zhang:1995tl,*Zhang:1997ui,Zhang:1997te}.

\section{Dimer correlation and structure factor}

\begin{center}
\begin{figure}
\includegraphics*[scale=0.30]{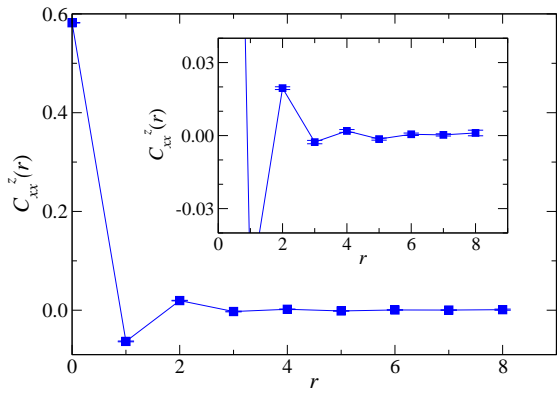}
\caption{Dimer-dimer correlation function $C^z_{xx}(\bfr)$ obtained on a $16\times 16$ lattice
at $U=6.8$. The correlation function is plotted along the $x$-axis.
The inset shows the behavior of $C^z_{xx}(\bfr)$ for $|\bfr|\geq 2$.}
\label{fig:dimer.corr}
\end{figure}
\end{center}

The dimer-dimer correlation function examined in this work is defined as
\be
  C^z_{\delta\delta}(\bfr)
  = \ob{ S^z_{\bfr+\hat\delta} S^z_\bfr S^z_{\hat\delta} S^z_0 } 
       -\ob{S^z_{\hat\delta} S^z_0}^2,
\ee
where $S^z_\bfr$ is the $z$-component spin $1/2$ operator at site $\bfr$.
$\delta=x$ or $y$ denotes the orientation of singlet bonds.
Typical behavior of the dimer-dimer correlation function is shown in Fig.~\ref{fig:dimer.corr}.
for a $16\times 16$ lattice at half-filling with interaction strength $U=6.8$. 

\begin{center}
\begin{figure}
\includegraphics*[scale=0.37]{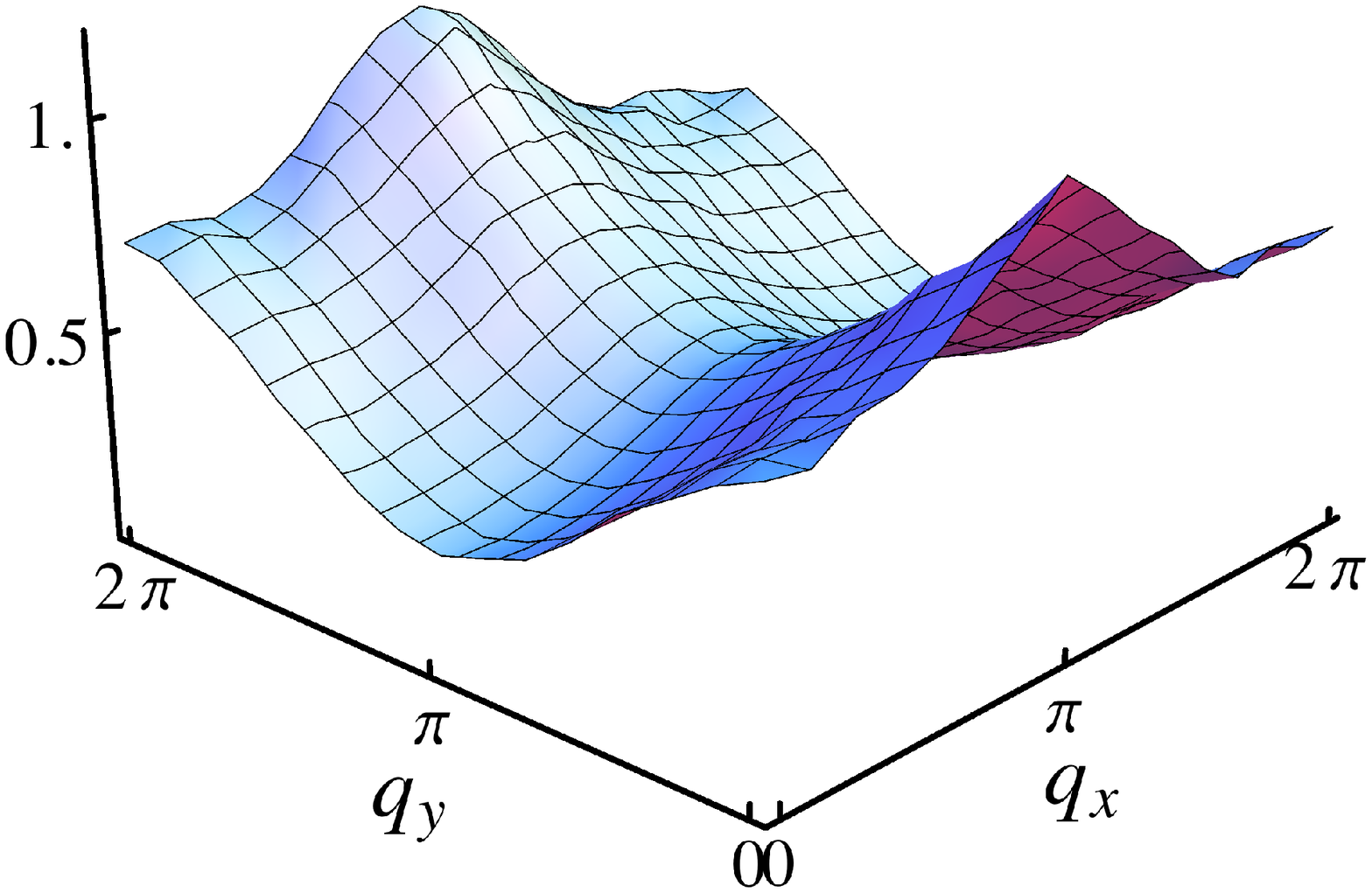}
\caption{Dimer structure factor $D_{xx}(\bfq,L)$ obtained on a $L=16$ square lattice
at half-filling and $U=6.8$.}
\label{fig:dimer.sk}
\end{figure}
\end{center}

We are also interested in the dimer structure factor. This quantity is defined as
\be
  D_{\delta\delta}(\bfq,L) =
  \frac 1 N \sum_\bfr\,e^{i\bfq\cdot\bfr}\,C^z_{\delta\delta}(\bfr).
\ee
Fig.~\ref{fig:dimer.sk} illustrates the dimer structure factor obtained
for a $L=16$ lattice at $U=6.8$. In this example, $D_{xx}(\bfq,16)$
peaks at the dimer characteristic wave vector $\bfq=(\pi,0)$, indicating a
weak dimer order on a finite lattice. To determine whether the ground state
is dimerized, we extrapolate $D_{\delta\delta}(\bfq,L)$ at $\bfq=(\pi,0)$ or
$(0,\pi)$ to the $L\rightarrow\infty$ limit, and extract its bulk value.
The results are shown in the main text.

\bibliography{fluxphase}